\documentclass[11pt,a4paper]{article}

\usepackage{layout}

\usepackage{graphicx}
\usepackage{epstopdf} % Need for eps figures in PDFLaTeX
\usepackage{amsmath}
\usepackage{amsfonts}
\usepackage{amssymb}
\usepackage{amsthm}

\numberwithin{equation}{section}

\usepackage{cite}
\usepackage{color}

\newcommand{\etal}{{\it et al.}}
\newcommand{\rmd}{{\mathrm d}}
\newcommand{\tr}{{\mathrm {tr}}}
\newcommand{\str}{{\mathrm {str}}}

\renewcommand{\Re}{{\mathrm {Re}}}
\renewcommand{\Im}{{\mathrm {Im}}}

\newcommand{\B}{{\mathbf B}}
\newcommand{\C}{{\mathbb C}}
\newcommand{\R}{{\mathbb R}}

\newcommand{\g}{{\mathbf g}}
\newcommand{\M}{{\mathbf M}}

\newcommand{\x}{{\mathbf x}}

\newtheorem{conjecture}{Conjecture}[section]
\newtheorem{definition}{Definition}[section]

\begin{document}

%%%%%%%%%%%%%%%%%%%%%%%%%%%%%
% TITLE PAGE & AUTHOR DETAILS
%%%%%%%%%%%%%%%%%%%%%%%%%%%%%
\title{Grassmann Variables and Pseudoclassical\\ Nuclear Magnetic Resonance}

\author{Robin A. Damion 
\\ 
\small{Nottingham, UK}\\
\\
%E-mail: r.a.damion@leeds.ac.uk
}

%\date{\today}

\maketitle

\begin{abstract}
The concept of a propagator is useful and is a well-known object in diffusion NMR experiments. Here, we investigate the related concept; the propagator for the magnetisation or the Green's function of the Torrey-Bloch equations. The magnetisation propagator is constructed by defining functions such as the Hamiltonian and Lagrangian and using these to define a path integral. It is shown that the equations of motion derived from the Lagrangian produce complex-valued trajectories (classical paths) and it is conjectured that the end-points of these trajectories are real-valued. The complex nature of the trajectories also suggests that the spin degrees of freedom are also encoded into the trajectories and this idea is explored by explicitly modeling the spin or precessing magnetisation by anticommuting Grassmann variables. A pseudoclassical Lagrangian is constructed by combining the diffusive (bosonic) Lagrangian with the Grassmann (fermionic) Lagrangian, and performing the path integral over the Grassmann variables recovers the original Lagrangian that was used in the construction of the propagator for the magnetisation. The trajectories of the pseudoclassical model also provide some insight into the nature of the end-points.    
\end{abstract}

\newpage
\tableofcontents

%%%%%%%%%%%%%%
% INTRODUCTION
%%%%%%%%%%%%%%
\section{Introduction}
Nuclear magnetic resonance (NMR) is a versatile experimental technique that has acquired a vast and varied collection of methods to measure what might broadly be categorised as molecular composition, structure, and dynamics. Within the last category are the translational motions of diffusing and flowing particles, and it is with these types of motion that we will be concerned with here. In particular, we will focus on the NMR measurement of translational motions of fluid particles~\cite{Callaghan,Price} by methods employing field gradients.

A common method of modeling NMR experiments of translational motion is via the propagator~\cite{KargerHeink}. The propagator is otherwise known as the conditional probability density~\cite{Wang} and is obtained as the Green's function to the underlying differential equation of the motion, usually a diffusion or Fokker-Planck equation~\cite{Risken} of some form. Once the propagator is determined it can be used to average the NMR phase factor $\exp(i\int\rmd t\gamma B(x,t))$ to obtain the time-dependence of the macroscopic magnetisation due to motion in the magnetic field $B(x,t)$.

In the context of diffusion or the Fokker-Planck equation, the propagator is an object which describes statistics of the stochastic process. For Markovian processes~\cite{Wang}, which will be assumed henceforth, the propagator together with the initial probability (marginal) density contain all the statistical information available. The propagator is thus an important function and there are many approaches available for its calculation. One such method is via the powerful technique of path integration~\cite{FeynmanHibbs,Kleinert,Wio} which was originally created to calculate the analogous quantity in quantum mechanics~\cite{FeynmanHibbs,Kleinert,Feynman,Schulman,Swanson}. In both these contexts, path integration is widely used, as it is in related subjects such as statistical physics. It is therefore surprising that it has been very rarely exploited in the context of NMR, with the exception of Le Doussal and Sen~\cite{LeDoussal}. 

There are possibly a few reasons for this absence of path integral methods. The first reason is probably because the propagator of the diffusion or Fokker-Planck equation is often more easily derivable by other methods in cases where the propagator is non-Gaussian, such as in the case of restricted or reflected diffusion~\cite{Grebenkov}, where an eigenfunction expansion, for example, works well. Path integral techniques deliver exact solutions in the case of Gaussian propagators but most of those cases are already known and derivable by other methods---for example, by cumulant expansions~\cite{Grebenkov}. Therefore, in regard to the propagator for the underlying stochastic process, unless there are any benefits of the perturbation methods particular to path integration, there might not be any circumstances in which path integration is advantageous in the context of NMR.

Suppose, however, that instead of averaging the NMR phase factor with the stochastic propagator, the propagator for the magnetisation is calculated directly. This is essentially what was done by Le Doussal and Sen~\cite{LeDoussal} who calculated the magnetisation propagator for diffusion in a time-independent parabolic magnetic field by analytically continuing the analogous quantum mechanical propagator for a particle in a quadratic potential~\cite{FeynmanHibbs,Schulman}.

It may seem that there is no significant difference between these two approaches, besides the fact that the propagator in the latter case becomes a complex function, but there is a subtlety which is not apparent until a path integral is formulated directly from the partial differential equation of the magnetisation. In the simplest case of diffusion in a linear field gradient, the term $i\gamma {\mathbf g}\cdot{\mathbf x}\tilde{M}(x,t)$ is added to the diffusion equation for the magnetisation $\tilde{M}(x,t)$ to form the Torrey-Bloch equation. It is this term which couples the diffusive motion with the spin (magnetic) degrees of freedom and introduces a non-trivial element to the dynamics. 

This non-trivial element is not apparent until one examines the analytical mechanics and the ensuing equations of motion via the Hamiltonian or Lagrangian. Usually, whether one considers classical or quantum mechanics, motion in a linear potential (in the absence of any damping terms such as viscosity) leads to accelerated motion. This is also true for diffusion NMR but it is found that the acceleration is \emph{imaginary} and thus the ``classical'' or ``particle'' trajectory given by the Hamiltonian or Lagrangian dynamics is generally complex-valued. 

This, at first, seems like an inconsistency: How can the position and velocity variables be complex? The answer is surely that we have unwittingly also encoded the spin degrees of freedom into the variables and this, in turn, leads to the intriguing possibility that we might also be able to include the spin degrees of freedom explicitly in the propagator. The results presented here are a preliminary investigation of this idea and related issues.  

In section~\ref{sec:path-integral} we begin by demonstrating the construction of a propagator for the magnetisation in a simple case of a diffusion NMR experiment. The propagator is constructed as a path integral over (paths of) the ordinary position variables but it is shown that the classical trajectory---the extremum path of the action---is complex-valued, and it is then conjectured that the end-points of these trajectories should be real-valued.
  
Section~\ref{sec:Grassmann} introduces the anticommuting Grassmann variables and shows that a simple Lagrangian produces equations of motion which mimic precessing magnetisation. A pseudoclassical Lagrangian (section~\ref{sec:pseudoclassical}) is then constructed from both the ordinary position variables and the Grassmann variables. An analysis of the equations of motion from this Lagrangian then lend some support to the notion that the end-points of the complex trajectories should be real. In sections~\ref{sec:grassmann-path-int}--\ref{sec:effective-bosonic-L} it is shown that by performing the path integration over the Grassmann variables the original propagator (involving just the position variables) is recovered.

In the final discussion, we mention some ways of extending the results. In particular, we discuss the possibility of incorporating radiofrequency fields into the propagator and---one of the main motivating factors behind this work---the possibility of constructing a supersymmetric model.

%%%%%%%%%%%%%%%%%%%%%%%%%%%%
% Path Integral
%%%%%%%%%%%%%%%%%%%%%%%%%%%%
\section{Path integral for an NMR propagator}\label{sec:path-integral}

%%%%%%%%%%%%%%%%%%%%%%%%
% Torrey-Bloch Equations
%%%%%%%%%%%%%%%%%%%%%%%%
\subsection{Torrey-Bloch equations}\label{sec:Torrey-Bloch}
The Bloch equations~\cite{Torrey,Slichter} describe the macroscopic magnetisation $\M(t)$ of an ensemble of spins in an applied magnetic field $\B(t)$
\begin{eqnarray}\label{eq:Bloch}
\frac{\rmd M_x}{\rmd t} &=& \gamma \left( \M\times \B\right )_x - \frac{M_x}{T_2},\nonumber \\
\frac{\rmd M_y}{\rmd t} &=& \gamma \left( \M\times \B\right )_y - \frac{M_y}{T_2},\\
\frac{\rmd M_z}{\rmd t} &=& \gamma \left( \M\times \B\right )_z - \frac{M_z-M_0}{T_1},\nonumber
\end{eqnarray}
where $\gamma$ is the gyromagnetic ratio, $T_1$ and $T_2$ are the spin-lattice and spin-spin relaxation times respectively, and $M_0$ is the equilibrium magnetization. In the absence of radiofrequency pulses, $\B(t)=B(t)\hat{\mathbf z}$ is purely polarized in the $z$-direction but might vary temporally (and spatially). In this case, the $z$-component of the magnetization is uninteresting (as far as our particular interest goes) and we can focus solely on the $x$ and $y$-components, so that
\begin{equation}\label{eq:Bloch-MxMy}
\frac{\rmd}{\rmd t}\begin{pmatrix} M_x  \\ M_y  \end{pmatrix} = 
\gamma B(t)\tilde{\varrho}_3\cdot \begin{pmatrix} M_x  \\ M_y  \end{pmatrix}
- \frac{1}{T_2}\begin{pmatrix} M_x  \\ M_y  \end{pmatrix}
\end{equation}
where 
\begin{equation}\label{eq:varrho-3}
\tilde{\varrho}_3 = \begin{pmatrix} 0 & 1  \\ -1 & 0  \end{pmatrix},
\end{equation}
which is the generator $\varrho_3$ of the Lie group SO(3), reduced to the $x,y$ components, \emph{i.e.} the generator of rotations about the $z$-axis.
 
More usually, a complex magnetisation is defined such that $\tilde{M}_{\pm}=M_x \pm iM_y$ and we can then transform the above vector equation into two decoupled components such that
\begin{equation}\label{eq:BlochMxy}
\frac{\rmd}{\rmd t}\begin{pmatrix} \tilde{M}_{+}  \\ \tilde{M}_{-}  \end{pmatrix} = 
-i\gamma B(t)\sigma_3\cdot \begin{pmatrix} \tilde{M}_{+}  \\ \tilde{M}_{-}  \end{pmatrix}
- \frac{1}{T_2}\begin{pmatrix} \tilde{M}_{+}  \\ \tilde{M}_{-}  \end{pmatrix}
\end{equation}
where 
\begin{equation}\label{eq:sigma-3}
\sigma_3 = \begin{pmatrix} 1 & 0  \\ 0 & -1  \end{pmatrix},
\end{equation}
is the $\mathfrak{su}(2)$ Pauli matrix corresponding to the $\mathfrak{so}(3)$ matrix $\varrho_3$. 

To generalize the differential equation~(\ref{eq:BlochMxy}) to include diffusive and coherent motion we can combine the above with the Fokker-Planck equation~\cite{Risken}
\begin{equation}\label{eq:Fokker-Planck}
\frac{\partial P}{\partial t} = {\mathbf \nabla}^2\left({\mathbf D} P\right) - {\mathbf \nabla} \left( {\mathbf v}P \right),
\end{equation}
where $P$ is a density distribution and, in the most general case, ${\mathbf D}$ is a temporally and spatially-dependent diffusion tensor, and $\mathbf v$ is a temporally and spatially-dependent velocity field.

We shall consider the less general case in which ${\mathbf D}=D$ is a temporally and spatially-independent scalar, in which case the Fokker-Planck equation reduces to the simpler form
\begin{equation}\label{eq:diffusion}
\frac{\partial P}{\partial t} = D{\mathbf \nabla}^2 P - {\mathbf \nabla} \left( {\mathbf v}P \right).
\end{equation}
The right-hand side of this partial differential equation can be combined with, say, the $\tilde{M}_{+}$ component (which will henceforth be denoted more simply as $\tilde{M}$) of equation~(\ref{eq:BlochMxy}) to form the Torrey-Bloch equation~\cite{Torrey,KuchelPages} with an advection term
\begin{equation}\label{eq:Torrey-Bloch}
\frac{\partial \tilde{M}}{\partial t} = D{\mathbf \nabla}^2 \tilde{M}  - {\mathbf \nabla} \left( {\mathbf v}\tilde{M} \right) -i\gamma B(t) \tilde{M} - \frac{\tilde{M}}{T_2},
\end{equation}
where it should also be mentioned that $T_2$ could, in principle, be spatially and temporally dependent.

Let us make one further simplification. The applied magnetic field $B(t)$ is composed of two components. The first is a large, time-independent, spatially-uniform magnetic field $B_0$. This is the field generated by, for example, a superconducting magnet which polarizes the individual spin magnetic moments and leads to a macroscopic magnetisation within the sample. During a typical diffusion measurement, linear magnetic field gradients are applied at various times which encode the motion of the sample molecules. These gradients may vary the $z$-component of the magnetic field linearly in any direction\footnote{Here $\x$ denotes a general position \emph{vector}. In the following sections, index notation will be used for convenience. There, the components of $\x$ will be denoted by $x^a$, $a=1,2,3$. This is equivalent to $\x = (x,y,z)^{\mathrm T}$.} $\x$ with strength $\g$. Thus the total magnetic field is given by
\begin{equation}\label{eq:total-B}
B(\x,t) = B_0 + \g (t)\cdot\x.  
\end{equation}
The constant $B_0$ can be neglected, however. This term merely modulates $\tilde{M}$ with the oscillating factor $\exp(-i\gamma B_0 t)$ and is not important in the dynamics that we will investigate below. Ignoring the $B_0$, we arrive at our defining equation for the magnetization $\tilde{M}$,
\begin{equation}\label{eq:M-PDE-v}
\frac{\partial \tilde{M}}{\partial t} = D{\mathbf \nabla}^2 \tilde{M}  - {\mathbf \nabla} \left( {\mathbf v}\tilde{M}\right) -i\gamma \g (t)\cdot\x \tilde{M} - \frac{\tilde{M}}{T_2}.
\end{equation}

In the next two sections, for simplicity, we will focus on the diffusive part of~(\ref{eq:M-PDE-v}), setting ${\mathbf v}=0$ and $R_2 = 1/T_2 = 0$. Thus,
\begin{equation}\label{eq:M-PDE}
\frac{\partial \tilde{M}}{\partial t} = D{\mathbf \nabla}^2 \tilde{M} -i\gamma \g (t)\cdot\x \tilde{M}
\end{equation}
will be our starting point. However, a more general differential equation is considered in Appendix~\ref{sec:Appendix-L-of-FP}.     

%%%%%%%%%%%%%%%%%%%%%%%%%%%%%%%%%%%%%%%%%%%%%%
% Hamiltonian, Lagrangian, Equations of Motion
%%%%%%%%%%%%%%%%%%%%%%%%%%%%%%%%%%%%%%%%%%%%%%
\subsection{Hamiltonian, Lagrangian, equations of motion}\label{sec:H-L-eom}
Clearly, the PDE for $\tilde{M}$, equation~(\ref{eq:M-PDE}) is not really an equation of energy. Nonetheless, it is analogous (to some extent) to a Schr\"{o}dinger equation and quantities such as the Hamiltonian and momentum operators may be defined analogously. 

To begin, we define the momentum operator as\footnote{As mentioned in the text, the diffusion or Fokker-Planck equation is a PDE for a stochastic process, not a quantum process, and therefore the usual definitions of quantities such as the Hamiltonian, Lagrangian, and momentum are slightly different. See Appendix~\ref{sec:Appendix-L-of-FP} for a more general derivation of the Lagrangian, Hamiltonian and associated definitions.}
\begin{equation}\label{eq:p-operator}
-ip_a = \frac{\partial }{\partial x^a}
\end{equation}
so that our phase-space coordinates are $\{x^a,p^a\}$ with $a=1,2,3$. The Hamiltonian, corresponding to the operator $-\partial /\partial t$, is then
\begin{equation}\label{eq:Hamiltonian}
H(x,p) = D p_a p^a + i g _a x^a,
\end{equation}
where index notation (and Einstein summation convention) is now being used, and where $\gamma$ has been absorbed into the definition of $\g$.

To define the Lagrangian we need an expression for the velocity $\dot{x}^a$. This is found using one of Hamilton's equations of motion (see Appendix~\ref{sec:Appendix-L-of-FP}), \emph{i.e.}
\begin{equation}\label{eq:x-dot}
\dot{x}_a = \frac{\partial H}{\partial (-ip^a)} = i\frac{\partial H}{\partial p^a} = i2Dp_a.
\end{equation}
The Lagrangian is then defined via the Legendre transform
\begin{eqnarray}\label{eq:Lagrangian}
L(x,\dot{x}) &=& -ip_a\dot{x}^a - H(x,p)\nonumber\\
		     &=& -\frac{1}{4D} \dot{x}_a\dot{x}^a - i g _a x^a.
\end{eqnarray}

Lagrange's equations of motion satisfy (in the absence of higher order time-derivatives than $\dot{x}$)
\begin{equation}\label{eq:L-eom}
\frac{\partial L}{\partial x^a} - \frac{\rmd}{\rmd t} \frac{\partial L}{\partial \dot{x}^a} =0,
\end{equation}
from which we obtain
\begin{eqnarray}
\ddot{x}^a &=& i2Dg^a(t),\label{eq:eom-x-ddot}\\
\dot{x}^a  &=& v^a + i2D\int_{t_0}^{t} \rmd ' g^a(t'),\label{eq:eom-x-dot}\\
x^a        &=& x^a_0 + v^a(t-t_0) + i2D\int_{t_0}^{t} \rmd t' \int_{t_0}^{t'} \rmd t'' g^a(t''),\label{eq:eom-x}
\end{eqnarray}
where $x^a_0$ and $v^a$ are constants of integration. In the case of constant $g$, the equations of motion simplify to (setting $t_0=0$)
\begin{eqnarray}
\ddot{x}^a &=& i2Dg^a,\label{eq:eom-x-ddot-const-g}\\
\dot{x}^a  &=& v^a + i2Dg^at,\label{eq:eom-x-dot-const-g}\\
x^a        &=& x^a_0 + v^at + iDg^at^2,\label{eq:eom-x-const-g}
\end{eqnarray}
where it is clearer that equation~(\ref{eq:eom-x-const-g}) for the trajectory in a constant magnetic field gradient $g^a$ has an imaginary acceleration and is, in general, complex-valued. Note that $x^a_0$ and $v^a$ are also generally complex numbers. By separating the real and imaginary parts for one component (one dimension), such that $x=\Re(x^a)$, $y=\Im(x^a)$ for fixed $a$, the trajectory can be made analogous to the two-dimensional parabolic motion of a body in a linear potential
\begin{eqnarray}
x        &=& x_0 + vt,\label{eq:eom-x-const-g-x1d}\\
y        &=& y_0 + ut + Dgt^2.\label{eq:eom-x-const-g-y1d}
\end{eqnarray}

The form of $y$ is also the same as that of the NMR phase as a particle moves at constant velocity $v$ in the direction of the gradient,
\begin{eqnarray}
\phi &=& \phi_0 + \int_0^t \rmd t' gx(t'),\nonumber\\
     &=& \phi_0 + gx_0t + \frac{1}{2}gvt^2,\label{eq:trajectory-phase} 
\end{eqnarray}
but there appears to be no obvious link between the imaginary component $y$ and the phase $\phi$ and, therefore, the imaginary part of the trajectory is not simply proportional to the phase. 

%%%%%%%%%%%%%%%%%%%%%%%%%%%%%%%%%%%%%%%%%%
% Path Integral, propagator, magnetisation
%%%%%%%%%%%%%%%%%%%%%%%%%%%%%%%%%%%%%%%%%%
\subsection{Path integral, propagator, magnetisation}\label{sec:S-path-integral}
With the Lagrangian defined in~(\ref{eq:Lagrangian}), the propagator (or Green's function) for the magnetisation of equation~(\ref{eq:M-PDE}) can be calculated using the technique of path integration~\cite{FeynmanHibbs, Feynman, Schulman, Kleinert}. Denoting the propagator by $G(x_T,T;x_0,t_0)$,
\begin{equation}\label{eq:path-integral}
G\left(x_T,T;x_0,t_0\right) = \int_{x_0,t_0}^{x_T,T} {\mathcal D} x\, e^S,
\end{equation}
where $\int {\mathcal D}x$ represents an infinite-dimensional integral over all \emph{paths} $x(t)$ beginning at $x_0, t_0$ and ending at $x_T,T$. The action $S$ is given by the integral of the Lagrangian;
\begin{equation}\label{eq:action-integral}
S = \int_{t_0}^{T} \rmd t L\left( x,\dot{x}\right).
\end{equation}

Since the Lagrangian~(\ref{eq:Lagrangian}) is quadratic in $\dot{x}$ and $x$, the path integral can be calculated exactly. The action is expanded in a variational manner
\begin{equation}\label{S-expansion}
S = S_0 + \delta S + \delta^2 S,
\end{equation}
where, because the Lagrangian is quadratic the variation terminates at the $\delta^2$ term. The first term $S_0$ is the action evaluated along the path given by the equations of motion, equation~(\ref{eq:eom-x}). In the context of quantum mechanics this would be the classical trajectory. By design, the first variation $\delta S=0$ due to the Euler-Lagrange equation~(\ref{eq:L-eom}). Finally, the third term $\delta^2 S$ is independent of $x$ and $\dot{x}$ and corresponds to the contribution to the path integral of the fluctuations around the classical trajectory. This term produces the normalisation to the propagator,
\begin{equation}\label{eq:path-integral-stage2}
G\left(x_T,T;x_0,t_0\right) =  \sqrt{\det \left( \frac{1}{2\pi}\frac{\partial^2 S_0}{\partial x_T \partial x_0} \right) } \exp\left(S_0(x_T,T,x_0,t_0)\right),
\end{equation}
and is the square-root of the Van Vleck determinant~\cite{Schulman} of the matrix $\partial^2 S_0/\partial x_T^a \partial x^b_0$.

Using the trajectory, equations~(\ref{eq:eom-x-dot}) and~(\ref{eq:eom-x}), $S_0$ can be evaluated as
\begin{eqnarray}\label{eq:S0}
S_0 &=& -\frac{(x_T-x_0)^2}{4D(T-t_0)}\nonumber\\ 
&-& ix_T^a\int_{t_0}^T \rmd t g_a(t)
+ i\frac{(x_T-x_0)^a}{T-t_0}\int_{t_0}^T \rmd t \int_{t_0}^{t} \rmd t' g_a(t')\nonumber\\  
&-& D\int_{t_0}^T \rmd t \left\{ \int_{t_0}^{t} \rmd t' g(t')\right\}^2
+ \frac{D}{T-t_0}\left\{\int_{t_0}^T \rmd t \int_{t_0}^{t} \rmd t' g(t')\right\}^2,
\end{eqnarray}
and the pre-exponential normalisation factor in~(\ref{eq:path-integral-stage2}) is
\begin{equation}\label{eq:normalisation-factor}
\sqrt{\det \left( \frac{1}{2\pi} \frac{\partial^2 S_0}{\partial x_T \partial x_0} \right)} = \frac{1}{\left( 4\pi D (T-t_0)\right)^{3/2}}.
\end{equation}
Note that in most NMR diffusion experiments the effective time-dependence of the gradient is chosen such that its integral at the time of measurement is zero. Thus, usually the second term on the right-hand side can be neglected.

In the case that $\g$ is constant, the propagator is given by
\begin{equation}\label{eq:G-const-g}
G\left(x_T,T;x_0,t_0\right) = \frac{1}{\left( 4\pi D (T-t_0)\right)^{3/2}} e^{ -\frac{(x_T-x_0)^2}{4D(T-t_0)} -\frac{i}{2}(T-t_0)g_a(x_T+x_0)^a -\frac{D}{12}g^2(T-t_0)^3 },
\end{equation} 
which compares analogously with the quantum mechanical propagator for a particle in a constant linear electric field~\cite{FeynmanHibbs,Schulman} and more exactly with the results of Stoller {\it et al.}~\cite{Stoller} and Le Doussal and Sen~\cite{LeDoussal}. However, the position variables in these references are real whereas, apparently, our variables $x^a_T$ and $x^a_0$ are complex. Thus, we see that if we arbitrarily set the end-points of the trajectories to be real numbers (whilst still allowing the rest of the trajectories to be generally complex-valued) our propagator agrees with previous results. 

At this stage, it is not clear whether this real constraint on the end-points is necessary and, if so, whether it may be specified less arbitrarily than by fiat. Conversely, if it is not necessary, it suggests that the imaginary part could be measurable experimentally. For now, however, we shall form the following conjecture:

\begin{conjecture}[Reality Constraint]\label{conj:reality-constraint}
Let the path ${\mathbf x}(t):[0,T] \rightarrow \C^3$ be parameterised by $t\in [0,T]$. Then, to obtain a valid propagator for the diffusion NMR experiment, it is a requirement that $\Im \{{\mathbf x}(0)\} = \Im \{{\mathbf x}(T)\} =0$. 
\end{conjecture}

To complete the picture before we move on into less well-known territory, let us compute the attenuation of magnetisation that occurs in this case; an NMR diffusion experiment for free diffusion in a linear, time-dependent, field gradient (which includes a time-dependence of the direction). For convenience, and as is usually the case, we will assume that the time-integral of the gradient over the time $T$ from initial excitation, $t_0=0$, to measurement, $t=T$, is zero. The magnetisation at $t=T$ is then given by
\begin{equation}\label{eq:M-Delta}
\langle \tilde{M}(T) \rangle =  \int_{-\infty}^{\infty} \rmd^3 x_T \int_{-\infty}^{\infty} \rmd^3 x_0 G\left(x_T,T;x_0,0\right)\tilde{M}(x_0,0).
\end{equation}

With the time-integral constraint on $\g(t)$, the action given above in equation~(\ref{eq:S0}) is only dependent on the difference $x_T-x_0$. Thus, by changing variables from $x_T$ to $r=x_T-x_0$, the two integrals can be separated;   
\begin{equation}\label{eq:M-T-stage2}
\langle \tilde{M}(T) \rangle =  \int_{-\infty}^{\infty} \rmd^3 r G\left(r,T\right) \int_{-\infty}^{\infty} \rmd^3 x_0 \tilde{M}(x_0,0).
\end{equation}
Letting the integral over $x_0$ be denoted by $M_0$ (which may be complex), the integrations over $r$ can be readily performed (being merely Gaussian integrals) to give
\begin{equation}\label{eq:Stejskal-Tanner}
\langle \tilde{M}(T) \rangle =  M_0 \exp\left( -D\int_{0}^{T} \rmd t \left\{ \int_{0}^{t} \rmd t' g(t')\right\}^2 \right),
\end{equation}
as was derived by Karlicek and Lowe~\cite{KarlicekLowe} and initially (in slightly less general form) by Stejskal and Tanner~\cite{StejskalTanner} by very different methods (also see Kuchel \etal~\cite{KuchelPages}). 

Thus, we see that the usual expression for the attenuation of the magnetisation in a diffusion experiment is derivable from a path integral method, and this provides us with some evidence that the method is correct. In fact, propagators can also be derived in more general situations, such as with diffusion tensors, velocity fields, potentials, and non-constant relaxation terms (see Appendix~\ref{sec:Appendix-L-of-FP} for a more general method of forming the path integral) but, except in special cases, the path integral is often not exactly calculable when the Lagrangian involves higher powers than the second.

Here, however, we shall not be concerned with these more general cases. Instead, the intention is to investigate the possibility of explicitly including the spin degrees of freedom into the path integral in such a way that they account for the imaginary part of the classical trajectory.   

%%%%%%%%%%%%%%%%%%%%%%%%%%%%%%
% Grassmann variables and spin
%%%%%%%%%%%%%%%%%%%%%%%%%%%%%%
\section{Grassmann variables and spin}\label{sec:Grassmann}
The imaginary parts of the trajectories, or classical paths, $x^a(t)$ given in section~\ref{sec:H-L-eom} seem to be related somehow to the spin degrees of freedom, or the precessing magnetisation. If this is true, we should be able to model the spin components explicitly in some form and, in doing so, this might reveal the nature of the imaginary part of the paths. Non-relativistic spins are usually described by Pauli matrices and their $\mathfrak{su}(2)$ algebra. However, incorporating Pauli matrices into path integrals is not straightforward and, besides, it is not clear that modeling the spin components quantum-mechanically is the correct way to proceed; after all, the observed magnetisation is of a macroscopic nature.

An alternative is to use Grassmann variables~\cite{Kleinert,Freund,Frappat}. These are often easier to handle within path integrals despite their anticommuting properties and can be used to model fermionic fields and spin at the classical level~\cite{BerezinMarinov1975,BerezinMarinov,Casalbuoni}. Scholtz~\cite{Scholtz} has also shown that it is possible to relate path integrals of Grassmann variables to bosonic path integrals over representation variables of $\mathfrak{su}(2)$. Here, we shall consider only Grassmann variables. 

%%%%%%%%%%%%%%%%%%%%%%%
% The Grassmann algebra
%%%%%%%%%%%%%%%%%%%%%%%
\subsection{The Grassmann algebra}\label{sec:Grassmann-algebra}
The real Grassmann algebra $\Gamma(n)$ of order $n$ is the algebra over $\mathbb R$ generated from the unit element $1$ and the $n$ Grassmann variables $\theta^i$, $i=1\dots n$, which satisfy the anticommutation relations
\begin{equation}\label{eq:anticommuator}
\{ \theta^i,\theta^j \} = \theta^i\theta^j + \theta^j\theta^i = 0.
\end{equation}
In particular, the anticommutation means that the variables are nilpotent,
\begin{equation}\label{eq:nilpotent}
\theta^i \theta^i = -\theta^i \theta^i = 0,
\end{equation}
and this implies that only a finite number, $2^n$, of monomials are possible; this is the dimension of the algebra.

Specialising to the case $n=3$, which is the case we will be interested in, $\dim \Gamma(3) =8$. The full set of monomials are
\begin{equation}
\{1\}, \{\theta^1,\theta^2,\theta^3\}, \{\theta^1\theta^2,\theta^2\theta^3,\theta^3\theta^1\}, \{\theta^1\theta^2\theta^3\} 
\end{equation}
and they partition the algebra into two parts; even and odd, or (in terms of a mod 2 grading) degree $\bar{0}$ and $\bar{1}$ respectively. Thus, $\Gamma(n) = \Gamma(n)_{\bar{0}} \oplus \Gamma(n)_{\bar{1}}$. Specifically, $\deg 1 = \deg \theta^i\theta^j = \bar{0}$ and  $\deg \theta^i = \deg \theta^i\theta^j\theta^k = \bar{1}$.

Further properties of Grassmann variables---particularly in regard to integration---can be found in Appendix~\ref{sec:Appendix-Berezin-integration}.

%%%%%%%%%%%%%%%%%%%%%%%%
% A Grassmann Lagrangian
%%%%%%%%%%%%%%%%%%%%%%%%
\subsection{A Grassmann Lagrangian}\label{sec:Grassmann-Lagrangian}
Now we shall posit a Lagrangian using the Grassmann algebra $\Gamma(3)$. Let~\cite{Junker,Kleinert,Freund}
\begin{equation}\label{eq:L-Grassmann}
L(\theta,\dot{\theta}) = \theta_i\dot{\theta}^i - B^i(t)\epsilon_{ijk} \theta^j\theta^k,
\end{equation}
where $B^i(t)$ is a time-dependent vector and $\epsilon_{ijk}$ is the antisymmetric Levi-Civita symbol;
\begin{equation}\label{eq:Levi-Civita}
\epsilon_{ijk} =
\begin{cases}
+1, & \mbox{if $\{i,j,k\}$ is an even permutation of $\{1,2,3\}$}\\ 
-1, & \mbox{if $\{i,j,k\}$ is an odd permutation of $\{1,2,3\}$}\\
0, & \mbox{if any indices are repeated.}
\end{cases}
\end{equation}

Using the Euler-Lagrange equation~(\ref{eq:L-eom}), and being careful with the commutation of odd-degree quantities, such as $\theta^i$ and $\partial /\partial \theta^i$, we find
\begin{equation}\label{eq:theta-Bloch}
\dot{\theta}^m = -\epsilon_{ij}^{\enspace m}B^i(t)\theta^j.
\end{equation}
By expanding the components, it is easy to see that this is informally equivalent to $\dot{\boldsymbol\theta}={\boldsymbol\theta}\times \B(t)$ and thus the $\theta$ variables precess around the vector $\B(t)$ just as with the Bloch equations given in~(\ref{eq:Bloch}). 

If we restrict $\B(t)$ to be fixed in direction such that $\B(t)=(0,0,B(t))$ the solution to the above differential equation is
\begin{eqnarray}
\theta^3 &=& \theta^3_0, \label{eq:theta3-eom}\\
\psi(t) &=& \psi_0 \exp\left( -i\int_0^t \rmd \tau B(\tau) \right), \label{eq:chi-eom}
\end{eqnarray}
where we have introduced the complex odd Grassmann variables $\psi = \theta^1 + i\theta^2$ and $\bar{\psi} = \theta^1 - i\theta^2$. 

Following Kleinert~\cite{Kleinert}, an even-degree quantity can be defined by
\begin{equation}\label{eq:s-def}
I_i = \frac{1}{4}\epsilon_{ijk} \theta^j\theta^k,
\end{equation}         
and, with this, the interaction part of the Lagrangian above becomes
\begin{equation}\label{eq:I-B}
B^i\epsilon_{ijk} \theta^j\theta^k \approx  I_iB^i = {\mathbf I}\cdot \B,
\end{equation}
which is analogous to the Zeeman interaction of the total spin angular momentum ${\mathbf I}$ with a magnetic field $\B$. The equations of motion for ${\mathbf I}$ corresponding to those of~(\ref{eq:theta3-eom}) and~(\ref{eq:chi-eom}) are
\begin{eqnarray}
I_3 &=& \frac{1}{2}\theta^1\theta^2 = \frac{1}{2}\theta^1_0\theta^2_0 = \frac{i}{4}\psi_0\bar{\psi}_0,\label{eq:I3-eom}\\
I_{+}(t) &=& \frac{i}{2}\theta^3_0\psi(t),\label{eq:Iplus-eom}
\end{eqnarray}
where we have defined $I_{+}=I_1+iI_2$. Note that whilst the first of the above two equations looks incorrect, it can be easily verified by the fact that $\rmd (\theta^1\theta^2)/\rmd t=0$ using the equations of motion~(\ref{eq:theta-Bloch}) with $B^i=B^3\delta^i_3$. Thus, $I_3$ is constant and proportional to the amplitudes of the precessing motion of $\theta^1$ and $\theta^2$.

Note that if the Grassmann variables $\theta^1,\theta^2,\theta^3$ or the quantities $I_1,I_2,I_3$ are regarded as vectors or states, they can be rotated amongst themselves by defining rotation operators
\begin{equation}\label{eq:R-operators}
R_a = -i\epsilon_{abc}\theta^b\frac{\partial}{\partial\theta_c},
\end{equation}
whose commutation relations form an algebra which is homomorphic to the $\mathfrak{su}(2)$ Lie algebra,  
\begin{equation}\label{eq:R-algebra}
[ R_a, R_b ] = i\epsilon_{ab}^{\ \ c} R_c.
\end{equation}

%%%%%%%%%%%%%%%%%%%%%%%%%%%
% Pseudoclassical mechanics
%%%%%%%%%%%%%%%%%%%%%%%%%%%
\section{Pseudoclassical mechanics}\label{sec:pseudoclassical}
Having established that we can model the spin degrees of freedom (or something akin to the precessing magnetisation) via a triplet of (fermionic) Grassmann variables $\theta^i$, we now ask whether it is possible to combine the Grassmann variables with the (bosonic) spatial variables $x^a$ to obtain equations of motion similar to those obtained in the original analysis, equations~(\ref{eq:eom-x-ddot})--(\ref{eq:eom-x}), or~(\ref{eq:eom-x-ddot-const-g})--(\ref{eq:eom-x-const-g}) in the case of a time-independent magnetic field gradient.

We shall attempt to simply combine the Lagrangians for a free particle, $L\approx -\dot{x}_a\dot{x}^a$, with the Lagrangian in~(\ref{eq:L-Grassmann}) for the Grassmann variables by adding them with an extra parameter $i\alpha$;
\begin{equation}\label{eq:L-combined}
L(x,\dot{x},\theta,\dot{\theta}) = -\frac{1}{4D} \dot{x}_a\dot{x}^a + i\alpha\theta_i\dot{\theta}^i - i\alpha B^i(x,t)\epsilon_{ijk} \theta^j\theta^k,
\end{equation}
where we have allowed $\B$ to be explicitly spatially-dependent so that the two sets of variables $x^a$ and $\theta^i$ will be coupled.

The equations of motion are
\begin{eqnarray}
\ddot{x}_a &=& i\alpha 2D \frac{\partial B^i}{\partial x^a}\epsilon_{ijk}\theta^j\theta^k,\label{eq:combined-x-ddot}\\
\dot{\theta}_m &=& -\epsilon_{ijm}B^i\theta^j.\label{eq:combined-theta-dot}
\end{eqnarray}
As previously, let us specify the magnetic field to be polarized in the $z$-direction, \emph{i.e.}, $\B=(0,0,B^3)$. This fixes the index $i=3$ in~(\ref{eq:combined-x-ddot}) and~(\ref{eq:combined-theta-dot}). We then specify $B^3(x,t)=B(x,t)=g_b(t)x^b$ to be linearly varying in space. Then
\begin{equation}
\frac{\partial B^i}{\partial x^a} = \frac{\partial B}{\partial x^a} \delta_3^i= g_a\delta_3^i. \label{eq:g-delta}
\end{equation}
Inserting this into the equations of motion, we see (as before) that $\theta^3$ decouples and we have
\begin{eqnarray}
\ddot{x}_a &=& i\alpha 4D g_a\theta^1_0\theta^2_0,\label{eq:combined-x-ddot-stage2}\\
\dot{\theta}_{i'} &=& g_ax^a\epsilon_{i'j'}\theta^{j'},\label{eq:combined-theta12-dot}\\
\dot{\theta}_3 &=& 0,\label{eq:combined-theta3-dot}
\end{eqnarray}
where $\epsilon_{i'j'}$ is the Levi-Civita symbol on two indices such that $\epsilon_{12}=1=-\epsilon_{21}$, $\epsilon_{11}=\epsilon_{22}=0$, and the indices $i'$ and $j'$ are reduced indices taking values $1$ or $2$ only. We have also used the result that $\theta^1\theta^2=\theta^1_0\theta^2_0$, a time-independent quantity.

The solution for the Grassmann variables is very similar to that in the previous section:    
\begin{eqnarray}
\theta^3 &=& \theta^3_0, \label{eq:combined-theta3-eom}\\
\psi(t) &=& \psi_0 \exp\left( -i\int_0^t \rmd \tau g_a(\tau) x^a(\tau) \right), \label{eq:combined-psi-eom}
\end{eqnarray}
where the exact time-dependence of $\psi(t)$ is dependent on the trajectory for $x^a(t)$, which is given by the solution to
\begin{equation}
\ddot{x}^a = i\alpha 8D g^a(t) I_3.\label{eq:combined-x-ddot-stage3}
\end{equation}

Comparing this equation of motion for $x^a$ with that of~(\ref{eq:eom-x-ddot}), it is tempting to suggest that $\alpha I_3= 1/4$. However, this notion overlooks the fact that $I_3$ is an even Grassmann parameter. Therefore, this suggestion does not make sense unless $\alpha$ can be regarded as an operator of some appropriate form. Despite this, if we do entertain the possibility that $\alpha$ and $I_3$ can be regarded as real numbers, and that the $I_i$ represent components of a spin angular momentum vector, then for spin-$\tfrac{1}{2}$ we expect that $I_3$ should give the component of the spin in the $z$-direction, and thus $I_3=1/2$. If this is correct, $\alpha=1/2$, which is the usual value of the coupling constant found in pseudoclassical~\cite{Freund} and supersymmetric models~\cite{Freund,Kleinert}. Nonetheless, this subtlety requires clarification. Although an answer to this possibly resides in the work of Scholtz~\cite{Scholtz}, it will not be pursued here.

Instead, if we take the above equation of motion at face value, it implies that the trajectory for $x^a(t)$ develops a component which is not only imaginary but also Grassmann-valued. Following Junker and Matthiesen~\cite{Junker1994,Junker1995}, and Junker~\cite{Junker}, we can thus split the trajectory into a so-called quasi-classical path $\breve{x}^a(t)$ and a Grassmann-valued part,
\begin{equation}\label{eq:quasi-classical}
x^a(t) = \breve{x}^a(t) + iq^a(t) I_3,
\end{equation}
where both $\breve{x}^a(t)$ and $q^a(t)$ are real functions of time. The quasi-classical path is just the trajectory due to the diffusive part of the motion in the absence of any coupling to the spin degrees of freedom and is given by the real solutions to $\ddot{\breve{x}}^a = 0$, and corresponds to the trajectory in~(\ref{eq:eom-x-const-g-x1d}).

Recalling the definition of $I_3$ from~(\ref{eq:I3-eom}), and inserting~(\ref{eq:quasi-classical}) into the trajectory~(\ref{eq:combined-psi-eom}) for $\psi(t)$,
\begin{align}\label{eq:psi-quasi-classical}
\psi(t) &= \psi_0 \exp\left( -i\int_0^t \rmd \tau g_a(\tau) \left[ \breve{x}^a(\tau) - \frac{1}{4}q^a(\tau) \psi_0\bar{\psi}_0 \right]\right) \nonumber\\ 
&= \psi_0 \left( 1 + \frac{i}{4}\int_0^t \rmd \tau g_a(\tau)q^a(\tau) \psi_0\bar{\psi}_0 \right) \exp\left( -i\int_0^t \rmd \tau g_a(\tau)\breve{x}^a(\tau)\right)  \nonumber\\ 
&= \psi_0 \exp\left( -i\int_0^t \rmd \tau g_a(\tau)\breve{x}^a(\tau)\right),
\end{align}
where we have used the properties of Grassmann variables to expand the exponential and finally eliminate the $q^a(t)$-term completely using $(\psi_0)^2=0$. Thus, we see that the trajectory for $\psi (t)$ is only dependent on the quasi-classical trajectory $\breve{x}^a(t)$ which, being a real function of time, implies that $\psi$ evolves only via its phase, which corresponds exactly to the phase as given by~(\ref{eq:trajectory-phase}).

The observation that the trajectory for $\psi(t)$ and the Grassmann part of the action is independent of $q^a(t)$ has a further consequence because we can clearly add an arbitrary function $\lambda^a(t)$ to $q^a(t)$ without changing anything (with regard to the Grassmann variables). However, there is a constraint on $\lambda^a(t)$. From~(\ref{eq:quasi-classical}) and~(\ref{eq:combined-x-ddot-stage3}) we also have 
\begin{equation}\label{eq:q-ddot}
\ddot{q}^a(t) I_3 = \alpha 8D g^a(t) I_3,
\end{equation}
from which $\ddot{\lambda}^a = 0$, and therefore
\begin{equation}\label{eq:lambda}
\lambda^a(t) = \lambda^a_0 + \nu^at, 
\end{equation}
for arbitrary constants $\lambda^a_0$ and $\nu^a$ or, alternatively, constant end-points of the trajectory for $\lambda^a(t)$: $\lambda^a_0$ and $\lambda^a_T$, so that $\nu^a = (\lambda^a_T-\lambda^a_0)/T$. 

These gauge-like degrees of freedom coincide exactly with the troublesome imaginary parts of the trajectory for $x^a(t)$ that was the subject of Conjecture~\ref{conj:reality-constraint} and correspond to shifting the spatial coordinates $x^a_{T,0} \to x^a_{T,0} + i\lambda^a_{T,0}$. However, the propagator (for example, the simple propagator of~(\ref{eq:G-const-g})) is generally not invariant under this shift, and neither does it appear to change as a result of some physical gauge transformation. Therefore, it appears that in order to preserve observables, we must set $\lambda^a_{T,0} = 0$.    

%%%%%%%%%%%%%%%%%%%%%%%%%%%%
% Grassmann path integration
%%%%%%%%%%%%%%%%%%%%%%%%%%%%
\subsection{Grassmann path integration}\label{sec:grassmann-path-int}
The fact that we have almost recovered the original equations of motion~(\ref{eq:eom-x-ddot}) for $x^a$ suggests that our modeling of the spin degrees of freedom has a chance of being correct or, at least, sufficient. However, our work is not yet complete. The equations of motion~(\ref{eq:combined-x-ddot-stage3}) in the previous section are Grassmann-valued and therefore what remains is to perform a path integral over the Grassmann variables.

The path integral over the Grassmann variables is not quite as straightforward~\cite{BerezinMarinov,Swanson} as with the spatial (commuting) variables. In particular, note that if the equations of motion~(\ref{eq:combined-theta12-dot}) and~(\ref{eq:combined-theta3-dot}) are substituted in to the Lagrangian~(\ref{eq:L-combined}) the Grassmann part of $L$ vanishes. This would imply that the propagator of the Grassmann variables is a constant, independent of the initial and final values, which cannot be correct. The source of the issue is that the Lagrangian is somewhat ambiguous and despite appearing to be formulated in terms of configuration-space variables $\theta^a$, $\dot{\theta}^a$, it is really a phase-space formulation. This can be seen by deriving the momentum variable
\begin{equation}
\pi_a = -\frac{\partial L}{\partial \dot{\theta}^a} = i\alpha\theta_a,
\end{equation}
which highlights the ambiguity between the position variables $\theta_a$ and momentum variables $\pi_a$. The path integral over Grassmann variables therefore requires a slightly different approach to that over ordinary commuting variables. Different methods can be found, for example, in Berezin and Marinov~\cite{BerezinMarinov}, Swanson~\cite{Swanson}, Junker~\cite{Junker}, and Aouachria~\cite{Aouachria}.

Probably the simplest method of performing the Grassmann path integral is by using time-slicing. To achieve this, we will first express the Grassmann part of the Lagrangian~(\ref{eq:L-combined}) in terms of $\theta^3$, $\psi = \theta^1 + i\theta^2$, and $\bar{\psi} = \theta^1 - i\theta^2$,  
\begin{equation}\label{eq:L-theta-psi}
L(\theta^3,\dot{\theta}^3,\psi,\dot{\psi},\bar{\psi},\dot{\bar{\psi}}) = 
i\alpha\theta_3\dot{\theta^3} + i\frac{\alpha}{2}\psi\dot{\bar{\psi}} + i\frac{\alpha}{2}\bar{\psi}\dot{\psi} + \alpha B^3(x,t)\psi\bar{\psi},
\end{equation}
where we have again assumed $\B (x,t)$ to be polarized in the $z$-direction, $\B (x,t) = (0,0,B^3(x,t))$. The Grassmann path integral that we now wish to evaluate is
\begin{equation}\label{eq:Grassmann-path-integral}
G_N\iiint {\mathcal D} \theta^3 {\mathcal D} \psi {\mathcal D} \bar{\psi} \exp\left(\int_0^T\rmd t L(\theta^3,\dot{\theta}^3,\psi,\dot{\psi},\bar{\psi},\dot{\bar{\psi}}) \right),
\end{equation}
where $G_N$ is an undetermined normalisation constant. 

The equations of motion for this Lagrangian are, of course, just those given in~(\ref{eq:combined-theta3-eom}) and~(\ref{eq:combined-psi-eom}) and the integral over $\theta^3$ is trivial because of~(\ref{eq:combined-theta3-eom}). Thus, the corresponding term effectively vanishes. The effective Lagrangian now reduces to 
\begin{equation}\label{eq:L-psi}
L(\psi,\dot{\psi},\bar{\psi},\dot{\bar{\psi}}) = i\frac{\alpha}{2}\psi\dot{\bar{\psi}} + i\frac{\alpha}{2}\bar{\psi}\dot{\psi} + \alpha B^3(x,t)\psi\bar{\psi}.
\end{equation}
Combining the two kinetic terms using an integration-by-parts, 
\begin{eqnarray}\label{eq:Grassmann-action-1}
S &=& \int_0^T \rmd t L(\psi,\dot{\psi},\bar{\psi},\dot{\bar{\psi}}) \nonumber\\
&=& i\frac{\alpha}{2}\left( \psi_T\bar{\psi}_T - \psi_0\bar{\psi}_0 \right) 
+ i\alpha \int_0^T \rmd t \bar{\psi}\dot{\psi} -iB(x,t)\psi\bar{\psi},
\end{eqnarray} 
where we have dropped the superscript on $B(x,t)$, and where we can then simplify the action by a change of variables
\begin{equation}\label{eq:psi-chi}
\chi = \psi\exp\left( i\int_0^t \rmd \tau B(x(\tau),\tau)\right) 
= \psi\exp\left( i\beta(t)\right).
\end{equation}
This action is now
\begin{equation}\label{eq:Grassmann-action-chi}
S = i\frac{\alpha}{2}\left( \chi_T\bar{\chi}_T - \chi_0\bar{\chi}_0 \right) 
+ i\alpha \int_0^T \rmd t \bar{\chi}\dot{\chi}.
\end{equation}

In the time-slicing method, the time-integral is replaced with a discretised version by splitting the time interval $[0,T]$ into $N$ slices of duration $\Delta t = T/N$ such that  
\begin{equation}\label{eq:action-sum}
i\alpha \int_0^T \rmd t \bar{\chi}\dot{\chi} \approx i\alpha\sum_{n=0}^{N-1} \bar{\chi}_{n+1}\frac{\chi_{n+1} - \chi_{n}}{\Delta t}\Delta t.  
\end{equation}
Expanding this summation, except for the end-points $n=0$ and $n=N$, it is found that for a fixed index $n=m$ there are three terms:
\begin{equation}\label{eq:m-terms}
i\alpha\left( -\bar{\chi}_{m+1}\chi_{m} + \bar{\chi}_{m} \chi_{m} - \bar{\chi}_{m}\chi_{m-1} \right).  
\end{equation}
Using the usual properties of Grassmann variables and Berezin integration (Appendix~\ref{sec:Appendix-Berezin-integration}), it is then straightforward to integrate the exponential of these terms over $\chi_m$ and $\bar{\chi}_m$;
\begin{equation}\label{eq:m-terms-integrated}
\iint \rmd \chi_m \rmd \bar{\chi}_m\exp \left\{ i\alpha\left( -\bar{\chi}_{m+1}\chi_{m} + \bar{\chi}_{m} \chi_{m} - \bar{\chi}_{m}\chi_{m-1} \right) \right\} = i\alpha\exp \left\{ -i\alpha \bar{\chi}_{m+1}\chi_{m-1} \right\}.  
\end{equation}
Using this result for each time-point in the summation except the end-points, the $N-1$ integrations over the pair $\chi_n$, $\bar{\chi}_n$ give
\begin{equation}\label{eq:N-1-terms-integrated}
\left( i\alpha\right)^{N-1} \exp \left\{ i\alpha\left( \bar{\chi}_T\chi_T - \bar{\chi}_T\chi_0 \right)\right\},  
\end{equation}
where the subscript $N$ has been replaced by the continuous-time label $T$. Clearly, as $N\rightarrow \infty$ the term $( i\alpha )^{N-1}$ does not converge to a finite value and therefore some form of regularisation is required; this is the purpose of the constant $G_N$ in~(\ref{eq:Grassmann-path-integral}). Setting $G_N = G_0( i\alpha )^{-N}$ and combining~(\ref{eq:N-1-terms-integrated}) with the boundary terms of the action in~(\ref{eq:Grassmann-action-chi}), the path integral is
\begin{equation}\label{eq:rho-chi}
\rho\left( \chi_T, T \mid \chi_0, 0\right) 
= \frac{-i}{\alpha}G_0\exp \left( -i\alpha\left[ \bar{\chi}_T\chi_0 - \frac{1}{2} \bar{\chi}_T\chi_T - \frac{1}{2} \bar{\chi}_0\chi_0 \right] \right).
\end{equation}   
Using~(\ref{eq:psi-chi}) to express $\rho$ in terms of $\psi$, we finally have
\begin{equation}\label{eq:rho-psi}
\rho\left( \psi_T, T \mid \psi_0, 0\right) 
= \frac{-i}{\alpha}G_0\exp \left( -i\alpha\left[ \bar{\psi}_T\psi_0 e^{-i\beta(T)} - \frac{1}{2} \bar{\psi}_T\psi_T - \frac{1}{2} \bar{\psi}_0\psi_0 \right] \right),
\end{equation}
which agrees in form with Swanson's evaluation~\cite{Swanson}. 

%%%%%%%%%%%%%%%%%%%%%%%%%%%%%%%
% Fermionic density supermatrix
%%%%%%%%%%%%%%%%%%%%%%%%%%%%%%%
\subsection{Fermionic density supermatrix}\label{sec:density-smatrix}
Previously, when we calculated the purely bosonic path integral for the diffusion NMR experiment (section~\ref{sec:S-path-integral}), we interpreted the result $G(x_T,T;x_0,t_0)$ as a classical propagator. However, we will now interpret the result of the fermionic path integral $\rho\left( \psi_T, T \mid \psi_0, 0\right)$ as a statistical mechanical density matrix.

Since $\rho\left( \psi_T, T \mid \psi_0, 0\right)$ is constructed from the Grassmann elements $\psi$ and $\bar{\psi}$, the density matrix is actually a density \emph{super}matrix of even degree (see Appendix~\ref{sec:Appendix-supermatrices}). In the case of ordinary density matrices or functions of the form $\rho\left( x_T, T \mid x_0, 0\right)$, for $x_T, x_0\in\R$, the trace of $\rho$ is taken by summing or integrating over all states $x_T$ such that $x_0=x_T$. Thus,
\begin{equation}\label{eq:rho-x-trace}
\grave{\rho}\left( T\right) = \tr\left( \rho\right) = \int\rmd x_T\rho\left( x_T, T \mid x_T, 0\right).
\end{equation}

For a supermatrix, the natural operation is not a trace but a supertrace (see Appendix~\ref{sec:Appendix-supermatrices}). Since $\rho$ is an even-degree supermatrix, the supertrace has the form of the difference of two traces;
\begin{equation}\label{eq:rho-str}
\check{\rho} = \str\left( \rho\right) = \tr\left( \rho_{+}\right) - \tr\left( \rho_{-}\right),
\end{equation}
where
\begin{eqnarray}
\rho_{+} &=& \rho\left( \psi_T, T \mid \psi_T, 0\right), \\
\rho_{-} &=& \rho\left( \psi_T, T \mid -\psi_T, 0\right),
\end{eqnarray}
which is interpreted as the sum over all periodic, $\rho_{+}$, and antiperiodic, $\rho_{-}$, orbits.

Calculating $\rho_{\pm}$ from~(\ref{eq:rho-psi}),
\begin{equation}\label{eq:rho-pm-psi}
\rho_{\pm} = \frac{-i}{\alpha}G_0\exp \left( -i\alpha\bar{\psi}_T\psi_T\left[ \pm e^{-i\beta(T)} - 1 \right] \right),
\end{equation}
which is similar to the state sum model of fermions on a circle with gauge group U(1) obtained by Barrett \etal~\cite{Barrett}. Using the properties of Grassmann variables to expand the exponential,
\begin{equation}\label{eq:rho-pm-psi-expand}
\rho_{\pm} = \frac{-i}{\alpha}G_0\left( 1 - i\alpha\bar{\psi}_T\psi_T\left[ \pm e^{-i\beta(T)} - 1 \right] \right),
\end{equation}
and therefore,
\begin{equation}\label{eq:rhoplus-rhominus}
\rho_{+} - \rho_{-} = -2G_0\bar{\psi}_T\psi_T e^{-i\beta(T)}.
\end{equation}

The above expression for $\rho_{+} - \rho_{-}$ is possibly related to the off-diagonal terms in the density matrix for a spin-$\tfrac{1}{2}$ particle. Since we have assumed that the magnetisation is purely in the transverse plane, we could represent it by the Bloch vector $\mathbf{u}=(\cos\beta,\sin\beta,0)$. The density matrix in this two-state representation $(\mid +\rangle , \mid -\rangle )$ is 
\begin{equation}\label{eq:rho-spin-half}
\rho = \frac{1}{2}\left( \mathbf{1}  + \mathbf{u}\cdot\boldsymbol{\sigma}  \right) 
= \frac{1}{2}\begin{pmatrix} 1 & e^{i\beta}  \\ e^{-i\beta} & 1  \end{pmatrix},
\end{equation}
where $\boldsymbol{\sigma}$ are the Pauli matrices. However, whether or not this is merely a superficial resemblance is unclear.

Returning to the density matrix in~(\ref{eq:rhoplus-rhominus}), we compute the supertrace by integrating over the Grassmann variables;
\begin{equation}\label{eq:rho-check}
\check{\rho}(T) = \str\left( \rho\right) = \int\int\rmd\psi_T\rmd\bar{\psi}_T\left( \rho_{+} - \rho{-}\right) =  -2G_0e^{-i\beta (T)},
\end{equation}
from which it would be convenient to set $G_0 = -1/2$. 

%%%%%%%%%%%%%%%%%%%%%%%%%%%%%%
% Effective bosonic Lagrangian
%%%%%%%%%%%%%%%%%%%%%%%%%%%%%%
\subsection{Effective bosonic Lagrangian}\label{sec:effective-bosonic-L}
We began, in section~\ref{sec:pseudoclassical}, by proposing that the Lagrangian for the diffusion NMR experiment can be formed by combining the stochastic diffusion Lagrangian with the Grassmann Lagrangian, as in equation~(\ref{eq:L-combined}), and showed that the combined Lagrangian reproduces something close to the correct equations of motion. We then showed that it was possible to perform the path integral over the Grassmann variables, without placing any restrictions on the form of $B(x,t)$ except that it was polarized in a single direction, which we took to be the $z$-direction. We are now in a position to recombine the result of the Grassmann path integration, equation~(\ref{eq:rho-check}), with the remaining bosonic path integral over $x^a$. The path integral is now
\begin{eqnarray}\label{eq:x-path-integral}
&&\iiint {\mathcal D} x^1{\mathcal D} x^2{\mathcal D} x^3 \exp\left(\int_0^T \rmd t -\frac{1}{4D}\dot{x}_a\dot{x}^a \right) \check{\rho}(T) \nonumber\\ 
&& =
\iiint {\mathcal D} x^1{\mathcal D} x^2{\mathcal D} x^3 \exp\left(\int_0^T \rmd t -\frac{1}{4D}\dot{x}_a\dot{x}^a \right) \exp\left(-i\beta (T)\right),
\end{eqnarray}
and using~(\ref{eq:psi-chi}) to express $\beta (T)$ in its integral form, it is easy to see that the effective bosonic Lagrangian is given by
\begin{equation}\label{eq:effective-L}
L(x,\dot{x}) = -\frac{1}{4D}\dot{x}_a\dot{x}^a - iB^3(x,t).
\end{equation}
This is precisely the bosonic Lagrangian given in equation~(\ref{eq:Lagrangian}) of section~\ref{sec:H-L-eom} with the special form $B^3(x,t) = g_a(t)x^a$. 

Furthermore, we note that the actual form of the original, purely bosonic, Lagrangian---the real part of the Lagrangian---was not important. This means that more general bosonic Lagrangians, such as those with potentials, relaxation terms, and velocity fields are also possible. For example,
\begin{equation}\label{eq:effective-L-potentials}
L(x,\dot{x}) = -\frac{1}{4D}\left(\dot{x} - \tilde{v}(x,t)\right)_a \left(\dot{x} - \tilde{v}(x,t)\right)^a - \tilde{U}(x,t) - iB^3(x,t),
\end{equation}
where $\tilde{v}^a$ and $\tilde{U}$ are effective velocity and relaxation (or potential) terms which are defined in~(\ref{eq:A-tilde-v}) and~(\ref{eq:A-tilde-U}) of Appendix~\ref{sec:Appendix-L-of-FP}.

%%%%%%%%%%%%
% DISCUSSION
%%%%%%%%%%%%
\section{Discussion}\label{sec:discussion}
Although we have mostly concentrated on the simplest case---diffusion in a linear field gradient---we have shown that the method of path integration over ordinary commuting position variables (as was demonstrated in section~\ref{sec:path-integral}) is appropriate to model the magnetisation of a diffusion NMR experiment. Except for some generalisations which were not considered, such as non-constant diffusion tensors, the only uncertainty in this approach is in the observation that the classical trajectories for position variables $x^a(t)$ become complex-valued. This led to the formation of Conjecture~{\ref{conj:reality-constraint}} which specified that the end-points of the trajectory should be real numbers.

The presence of the complex trajectories suggested that there were some internal degrees of freedom which were associated with the precessing magnetisation of an NMR experiment, and to model these degrees of freedom explicitly we used the anticommuting Grassmann variables. Grassmann variables have previously been used many times to model spin at the classical level, with some of the earliest work being by Berezin and Marinov~\cite{BerezinMarinov1975,BerezinMarinov}, and Casalbuoni~\cite{Casalbuoni}, and can be considered to be the classical analogue of spin as $\hbar \rightarrow 0$. There are other ways to model spin or precessing magnetisation (such as with representation variables of SU(2)~\cite{Scholtz}, or as geometric algebra~\cite{Lasenby}, or based on the spherical top~\cite{Schulman1968}) but here, despite the unfamiliarity of such objects in NMR, we have employed Grassmann variables because of their relative simplicity and their connection with other fields of physics and mathematics.

By combining the commuting variables $x^a$ with the Grassmann variables $\theta^i$, and thus producing a pseudoclassical model, it was shown that parts of the imaginary component of the trajectory for $x^a(t)$ appeared to be of a purely internal nature. That is, the imaginary trajectory of the form $\lambda^a(t) = \lambda^a_0 + \nu^at$ has no effect on the trajectories of the Grassmann variables or their path integral and only serves to shift the end-points of the position variables such that $x^a_{0,T}\rightarrow x^a_{0,T} + i\lambda^a_{0,T}$. Since such a shift will, in general, affect observables (such as the average magnetisation) and there appears to be no obvious physical interpretation of the imaginary shift, we conclude that the end-points $x^a_{0,T}$ are purely real-valued, in agreement with Conjecture~{\ref{conj:reality-constraint}}.

There are a number of directions in which this work could be extended. As already mentioned, no attempt has been made to model the effects of boundary conditions such as reflective or absorbing walls, permeable membranes, etc. Unfortunately, even in fairly simple cases the propagators are generally difficult to solve analytically. Although the perturbation methods of path integration possibly offer some interesting ways of approaching these cases, techniques such as those of Mitra \etal~\cite{Mitra} and Ghadirian \etal~\cite{Ghadirian} might be more profitable.

In this paper, we modeled the transverse magnetisation only and implicitly assumed that any radiofrequency pulses merely modified the effective time-dependence of the magnetic field. In many cases, this simplification is sufficient but, for example, in the presence of inhomogeneous radiofrequency fields this approach might not be adequate. The work of Aouachria~\cite{Aouachria, Aouachria2011} has shown that radiofrequency fields can be straightforwardly included as potentials in the Grassmann variables and it would be interesting to explore the interaction between such potentials and the stochastic part of the Lagrangian. 

The pseudoclassical model involves a Lagrangian which is a combination of commuting variables $x^a$ and anticommuting variables $\theta^i$. Translating this into the parlance of particle physics, we have bosonic fields $x^a(t)$ and fermionic fields $\theta^i(t)$ on the one-dimensional manifold of time, and the natural question arises of whether there could exist a supersymmetry~\cite{Junker,Freund,Cooper1995} between these fields. There are some good reasons to think that there could be.

The Torrey-Bloch equation is somewhat similar to the Pauli equation for a spin-$\tfrac{1}{2}$ quantum particle in a magnetic field, which is known to exhibit a supersymmetry~\cite{Kleinert}. Whilst the Torrey-Bloch equation is not simply an analytical continuation (Wick rotation) of the Pauli equation, it is known that the diffusion or Fokker-Planck equation can also possess a supersymmetry~\cite{Andrea,Junker}. In the absence of a magnetic field, it is easy to show that the Lagrangian $-\dot{x}_a\dot{x}^a + i\theta_a\dot{\theta}^a$ has a trivial supersymmetry---trivial in the sense that the dynamics of the bosonic and fermionic fields are completely independent and therefore uninteresting. Adding the magnetic field so that the two sets of fields couple, as with the Lagrangian~(\ref{eq:L-combined}), however, destroys the supersymmetry. This is because supersymmetry requires that there is also some form of additional potential which is related to the magnetic field. In the case of the Pauli equation, the supersymmetry is a minimal $\mathcal{N}=1$ supersymmetry and the related potential is the vector potential of the magnetic field~\cite{Kleinert}. In the case of Witten-type~\cite{Witten1981,Witten1982,Junker} $\mathcal{N}=2$ supersymmetry (which we have not explored here) the relation is less clear in the present context and requires clarification.

It is fairly certain that supersymmetric Lagrangians can be constructed for both $\mathcal{N}=1$ and $\mathcal{N}=2$ supersymmetry but it is not yet clear whether these models correspond to physically realisable systems in the context of diffusion NMR. It is also not yet clear whether such supersymmetries have any theoretical or experimental consequences. This is the focus of current research.

Regardless of whether any form of supersymmetric model can be constructed the results presented here---for the ordinary (bosonic) path integral or for the pseudoclassical model---form the basis on which path-integral methods can be developed for diffusion NMR experiments.

%%%%%%%%%%%%%%%%%%
% ACKNOWLEDGEMENTS
%%%%%%%%%%%%%%%%%%
\section*{Acknowledgements}
My thanks are due to Erick Hinds Mingo for his work on understanding Grassmann path integration (section~\ref{sec:grassmann-path-int}). EHM was funded by EPSRC at the University of Leeds, School of Physics \& Astronomy, during the summer of 2014.

%%%%%%%%%%%%%%%%%%%%%%%%
% APPENDICES
%%%%%%%%%%%%%%%%%%%%%%%%
\appendix
%%%%%%%%%%%%%%%%%%%%%%%%
% Appendix A
%%%%%%%%%%%%%%%%%%%%%%%%   
\section{Appendix: Lagrangian of the Fokker-Planck equation}\label{sec:Appendix-L-of-FP}
Here is provided a derivation of the Lagrangian corresponding to the Fokker-Planck equation. In the process, we shall also discover the correct form of the Legendre transform and its associated equations of motion. The derivation follows closely that given by Wio~\cite{Wio}.

We begin with a fairly general Fokker-Planck equation of the form 
\begin{equation}\label{eq:A-FP}
\frac{\partial f}{\partial t} = {\mathbf \nabla}^2\left(D(x,t) f\right) - {\mathbf \nabla} \left( {\mathbf v}(x,t)f \right) - U(x,t)f,
\end{equation}
where $f=f(x,t)$, $D(x,t)$ is a scalar diffusion coefficient, ${\mathbf v}(x,t)$ is a velocity field, $U(x,t)$ is a sink or source term. In the context of diffusion NMR, $U(x,t)$ may include transverse relaxation and the imaginary NMR (Bloch) term $i\gamma B(x,t)$, as in equations~(\ref{eq:Torrey-Bloch}) or~(\ref{eq:M-PDE-v}). Although this Fokker-Planck equation is quite general, for simplicity, we do not consider the more general cases of tensor diffusion or the effects of topological constraints (such as impermeable barriers). Actually, the case of a constant diffusion tensor is straightforward but becomes more complicated if it is spatially or time-dependent.

Expanding the derivatives in~(\ref{eq:A-FP}), we can rewrite it as  
\begin{equation}\label{eq:A-FP-effective}
\frac{\partial f}{\partial t} = D(x,t)\nabla^2 f - \tilde{{\mathbf v}}(x,t)\cdot {\mathbf \nabla} f - \tilde{U}(x,t)f,
\end{equation}
where
\begin{gather}
\tilde{\mathbf v} = {\mathbf v} - 2{\mathbf \nabla} D,\label{eq:A-tilde-v}\\
\tilde{U}  = U + {\mathbf \nabla}\cdot {\mathbf v} - \nabla^2 D.\label{eq:A-tilde-U}
\end{gather}

Let $f(x_2,t)$ be a solution of~(\ref{eq:A-FP-effective}) at ${\mathbf x}={\mathbf x}_2$. Then the solution at an infinitesimal time increment $\epsilon$ later is
\begin{equation}\label{eq:f-t+epsilon}
f(x_2,t+\epsilon)\approx f(x_2,t) + \epsilon \frac{\partial}{\partial t}f(x_2,t).
\end{equation}
With this, we can write,
\begin{eqnarray}\label{eq:f-t+epsilon-x1}
f(x_2,t+\epsilon) &=& \left( 1 + \epsilon \frac{\partial}{\partial t} \right) f(x_2,t)\nonumber\\
&=&  \int\rmd^3 x_1 \left( 1 + \epsilon \frac{\partial}{\partial t} \right) \delta^3(x_1-x_2) f(x_1,t),
\end{eqnarray}
where a three-dimensional delta-function integral has been inserted to relate the position ${\mathbf x}_2$ at time $t+\epsilon$ to ${\mathbf x}_1$ at time $t$.

Using the Fokker-Planck equation~(\ref{eq:A-FP-effective}), we can replace the time-derivative in~(\ref{eq:f-t+epsilon-x1}) with a derivative operator which operates on ${\mathbf x}_2$. That is,
\begin{equation}\label{eq:f-FP-x2}
f(x_2,t+\epsilon) = 
\int\rmd^3 x_1 \left( 1 + \epsilon\left[ D(x_2,t)\nabla^2 - \tilde{{\mathbf v}}(x_2,t)\cdot {\mathbf \nabla} - \tilde{U}(x_2,t)  \right] \right) \delta^3(x_1-x_2) f(x_1,t),
\end{equation}
where $\nabla$ operates on $x_2$ only.

Replacing the delta-function with its Fourier representation,
\begin{equation}\label{eq:Fourier-delta}
\delta^3(x_1-x_2) = \int\frac{\rmd^3 p}{(2\pi)^3} e^{i{\mathbf p}\cdot ({\mathbf x}_1 - {\mathbf x}_2)},
\end{equation}
we can effectively replace ${\mathbf \nabla}$ with $-i{\mathbf p}$ in~(\ref{eq:f-FP-x2}) such that it becomes the phase-space integral
\begin{equation}\label{eq:f-FP-phase-space}
\int\rmd^3 x_1 \int\frac{\rmd^3 p}{(2\pi)^3} \left( 1 + \epsilon\left[ D(x_2,t)(-ip)^2 - \tilde{{\mathbf v}}(x_2,t)\cdot (-i{\mathbf p}) - \tilde{U}(x_2,t)  \right] \right) e^{i{\mathbf p}\cdot ({\mathbf x}_1 - {\mathbf x}_2)} f(x_1,t).
\end{equation}
Approximating the term in square brackets as an exponential, we can write this as
\begin{equation}\label{eq:f-FP-phase-space-exp}
\int\rmd^3 x_1 \int\frac{\rmd^3 p}{(2\pi)^3} \exp\left( \epsilon\left[ i{\mathbf p}\cdot \frac{({\mathbf x}_1 - {\mathbf x}_2)}{\epsilon} + D(x_2,t)(-ip)^2 - \tilde{{\mathbf v}}(x_2,t)\cdot (-i{\mathbf p}) - \tilde{U}(x_2,t)  \right] \right) f(x_1,t).
\end{equation}

We can now interpret the term in square brackets in~(\ref{eq:f-FP-phase-space-exp}) as the infinitesimal action integral 
\begin{equation}\label{eq:FP-action}
S = \int_t^{t+\epsilon}\rmd \tau L(x,\dot{x}) = \int_t^{t+\epsilon}\rmd \tau \left[ -i{\mathbf p}\cdot \dot{{\mathbf x}} + D(x,\tau)(-ip)^2 - \tilde{{\mathbf v}}(x,\tau)\cdot (-i{\mathbf p}) - \tilde{U}(x,\tau)  \right],
\end{equation}
where the term $({\mathbf x}_2 - {\mathbf x}_1)/\epsilon$ has been identified as the velocity $\dot{\mathbf x}$. From this it is easy to see that the Lagrangian can be inferred to be
\begin{equation}\label{eq:FP-Lagrangian}
\begin{split} 
L(x,\dot{x}) 
&= (-i{\mathbf p})\cdot \dot{{\mathbf x}} - \left[ -D(x,t)(-ip)^2 + \tilde{{\mathbf v}}(x,t)\cdot (-i{\mathbf p}) + \tilde{U}(x,t)  \right],\\
&= (-i{\mathbf p})\cdot \dot{{\mathbf x}} - H(x,(-ip)),
\end{split}
\end{equation}
where, since the first line has been deliberately written in the form of the Legendre transform, we can identify the Hamiltonian in the second line. Note that the Legendre transform above is equivalent to the usual form with ${\mathbf p}\mapsto -i{\mathbf p}$. From this, we can deduce that the Hamiltonian and Lagrangian equations of motion are
\begin{align}\label{eq:H-L-eoms}
\frac{\partial H}{\partial x^a} &= i\dot{p}_a, & \frac{\partial H}{\partial p_a} &= -i\dot{x}^a,\\
\frac{\partial L}{\partial x^a} &= -i\dot{p}_a, & \frac{\partial L}{\partial \dot{x}^a} &= -ip_a.
\end{align}

To complete the derivation so that $L$ is expressed in terms of ${\mathbf x}$ and $\dot{{\mathbf x}}$ only, we can use Hamilton's equations above to obtain
\begin{equation}
\frac{\partial H}{\partial p_a} = -i\dot{x}^a = 2D(x,t)p^a - i\tilde{v}^a, 
\end{equation}
and therefore,
\begin{equation}
 -ip^a = \frac{\tilde{v}^a - \dot{x}^a}{2D}.
\end{equation}
Substituting this into the Lagrangian~(\ref{eq:FP-Lagrangian}) we obtain the final result
\begin{equation}\label{eq:FP-Langrangian-final}
L(x,\dot{x}) = -\frac{\left( \dot{x} - \tilde{v}(x,t) \right)^2}{4D(x,t)} - \tilde{U}(x,t).
\end{equation}

%%%%%%%%%%%%%%%%%%%%%%%%
% Appendix B
%%%%%%%%%%%%%%%%%%%%%%%%   
\section{Appendix: Berezin integration}\label{sec:Appendix-Berezin-integration}
A brief introduction to the real Grassmann algebra was given in section~\ref{sec:Grassmann-algebra}. Here, we equally briefly explain integration over Grassmann variables, known as Berezin integration~\cite{Freund,BerezinMarinov,DeWitt}.

Let $\eta\in\Gamma(1)$ be a real Grassmann variable. Then, since $\eta^2=0$, any function of $\eta$, $f(\eta )$, can be expanded very simply as
\begin{equation}\label{eq:f-eta}
f(\eta ) = a + b\eta,  
\end{equation}
for constants $a$, $b$, which may in general be of even or odd degree. To perform an integral over $\eta$, we need the differential $\rmd \eta$. This is also an odd Grassmann element and thus
\begin{equation}\label{eq:eta-deta}
\eta \rmd \eta = -\rmd \eta\, \eta.
\end{equation}
This means that in defining the integral, a convention must be established. Here we use the following;
\begin{eqnarray}
\int \rmd\eta &=& 0,\\
\int \rmd\eta\, \eta &=& 1.
\end{eqnarray}
Note that Berezin integration is always indefinite. Applying the above to a function of $\eta$, we have 
\begin{equation}\label{eq:f-integral}
\int \rmd\eta f(\eta ) = \int \rmd\eta \left( a + b\eta \right) 
= (-1)^{\deg b}b\int \rmd\eta \eta = (-1)^{\deg b}b.  
\end{equation}

Extending these rules to integration involving more than one Grassmann variable is straightforward except that care must be taken with regard to the signs. Thus, for $\eta_1,\eta_2\in\Gamma(2)$,
\begin{eqnarray}
\int \rmd\eta_i &=& 0,\\
\int \rmd\eta_i \eta_j &=& \delta_{ij},\\
\iint \rmd\eta_1\eta_1 \rmd\eta_2\eta_2 &=& - \iint \rmd\eta_1\rmd\eta_2 \eta_1\eta_2 \nonumber\\
&=& \iint \rmd\eta_1\rmd\eta_2 \eta_2\eta_1 = 1.  
\end{eqnarray}
For a function of two Grassmann variables $f(\eta_1,\eta_2)$, which has the general expansion
\begin{equation}
f(\eta_1,\eta_2) = a + b_1\eta_1 + b_2\eta_2 + c_{12}\eta_1\eta_2,
\end{equation}
\begin{equation}
\iint \rmd\eta_1\rmd\eta_2 f(\eta_1,\eta_2) = \iint \rmd\eta_1\rmd\eta_2 c_{12}\eta_1\eta_2 = -c_{12},
\end{equation}
regardless of the degree of $c_{12}$.

For more subtleties, in particular for complex Grassmann variables, see, for example, Swanson~\cite{Swanson}.

%%%%%%%%%%%%%%%%%%%%%%%%
% Appendix C
%%%%%%%%%%%%%%%%%%%%%%%%     
\section{Appendix: Supermatrices and supertraces}\label{sec:Appendix-supermatrices}
\begin{definition}[Supermatrix~\cite{Frappat}]
A matrix $M$ is called a real (or complex) supermatrix if its entries have
values in a real (or complex) Grassmann algebra $\Gamma = \Gamma_{\bar{0}} \oplus \Gamma_{\bar{1}}$. More precisely, consider the set of $(m + n)\times (p + q)$ supermatrices $M$ of the form
\begin{equation}
M = \begin{pmatrix} A & B  \\ C & D  \end{pmatrix}
\end{equation}
where $A$, $B$, $C$, $D$ are $m \times p$, $m \times q$, $n \times p$ and $n \times q$ submatrices respectively. The supermatrix $M$ is called even (or of degree $\bar{0}$) if $A,D\in\Gamma_{\bar{0}}$ and $B,C\in\Gamma_{\bar{1}}$, while it is called odd (or of degree $\bar{1}$) if $A,D\in\Gamma_{\bar{1}}$ and $B,C\in\Gamma_{\bar{0}}$.
\end{definition} 

\begin{definition}[Supertrace~\cite{Frappat}]
The supertrace of the supermatrix $M$ is defined by
\begin{equation}
\str (M) = \tr (A) - (-1)^{\deg M}\tr (D) = 
\begin{cases} \tr (A) - \tr (D) & \mbox{if } M\in\Gamma_{\bar{0}} \\ 
              \tr (A) + \tr (D) & \mbox{if } M\in\Gamma_{\bar{1}}. 
\end{cases} 
\end{equation} 
\end{definition}

%%%%%%%%%%%%
% REFERENCES

\end{document}